\documentstyle[aps,prl,multicol,epsfig]{revtex}

\newcommand{\be}{\begin{equation}}
\newcommand{\ee}{\end{equation}}
\def\ket#1{|#1\rangle}
\def\bra#1{\langle #1 |}

\draft
\tighten

\begin{document}
\title{Transitions in quantum networks}

\author{P\"aivi T\"orm\"a}
\address{Institute for Theoretical Physics, University of Innsbruck,
Technikerstra{\ss}e 25, 6020 Innsbruck, Austria}
\maketitle

\begin{abstract}
We consider transitions in quantum networks analogous to 
those in the two-dimensional Ising model. We show that for a network 
of active components the transition is between the quantum and 
the classical behaviour of the network, and the critical amplification
coincides with the fundamental quantum cloning limit.
\end{abstract}

\pacs{03.65.-w, 03.67.-a, 05.50.+q, 42.50.-p}

\begin{multicols}{2}[]

Precise control over single quantum systems is essential in testing
and harnessing of quantum mechanics. This has become possible with the
advances in laser cooling and trapping techniques and manipulation of
optical elements in the one-photon level. The availability of single
quantum systems has fed the interest in {\it quantum networks}: A
quantum computer \cite{Qcomp} is a network of individual quantum
systems, where any two of the nodes can interact with each other. Most
quantum private communication schemes \cite{Qcomm} are networks of two
or three nodes. In addition to these information processing and
communication related applications, networks of optical components
\cite{Zeil1} and avoided crossings in multilevel systems
\cite{Bow,Harmin2} have been considered in order to study higher
dimensional quantum interference effects.

According to statistical physics, a set of probabilistically behaving
individual systems can exhibit critical behaviour when connected. In
this paper we consider the question whether transition phenomena
exist in networks of systems which behave probabilistically {\it not
because of finite temperature but due to their quantum nature};
transitions are known to exist in Ising quantum chain models.  We
define a model of a quantum network which carries in its structure a
formal analogy to the two-dimensional Ising model. Such networks can
be experimentally realized by various active (energy-consuming) or
passive (energy-preserving) components. It is found that transitions
do take place and we are able to give them a clear physical
interpretation.  For active systems the transition is between quantum
and classical, for passive systems between diabatic and adiabatic
behaviour of the network. The transition phenomenon is clearly
reflected in observable quantities.

The Ising model describes a set of two-state systems which interact
with their nearest neighbours; a quantum analogy of such a setup can
be experimentally realized in various ways, as will be explained
below.  Fig.1 shows schematically a 2-D quantum network with nearest
neighbour interactions. To define the building blocks of this network
we now take a closer look at the Ising model.

The two-state systems in the Ising model, let us say spins, are on a
2-D lattice of the size $N\times M$.  Since only nearest neighbour
interactions are taken into account, the total energy of the system
can be expressed using the energy $E(\mu)$ of one column (with the
spin configuration $\mu$) and the energy $E(\mu,\mu')$ between two
columns.  Let $s_k=\pm1$ denote the values of individual spins and
$\epsilon$ be the absolute value of the energy of a spin-spin
interaction. Then the energies can be written as $E(\mu) = - \epsilon
\sum_{k=1}^N s_k s_{k+1}$ and $E(\mu, \mu') = - \epsilon \sum_{k=1}^N
s_k {s'}_k$.
\narrowtext
\vbox{
\begin{figure}
\begin{center}
\epsfig{file=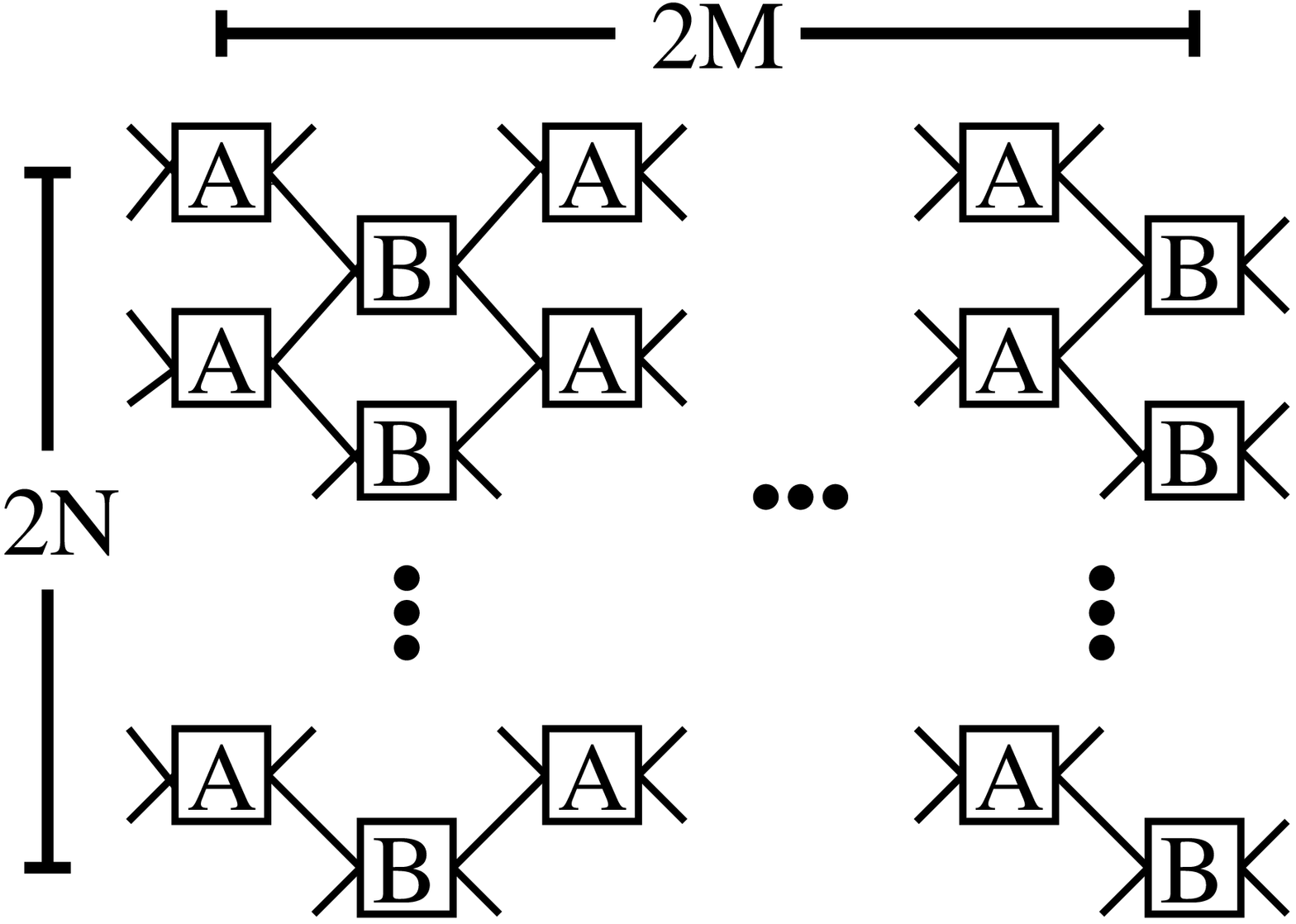,width=1.5in,angle=270}
\end{center}
\caption{A quantum network where the nodes are connected to their
nearest neighbours. The boxes \fbox{A} and \fbox{B} denote the
transformations performed at the nodes; when they are defined by
Eq.(1) the network has a relation to the Ising model.}
\end{figure}
} 
The partition function $Q(T)$ can be expressed in a simple form by
defining a $2^N\times 2^N$ matrix ${\cal P}$ whose matrix elements are
the thermal weight factors corresponding to a particular spin
configuration of two neighbouring columns $\bra{\mu}{\cal P}\ket{\mu'}
\equiv e^{-\beta [E(\mu,\mu')+E(\mu)]}$. With this notation $Q(T) = Tr
{\cal P}^M = \sum_{\alpha=1}^{2^N} (\lambda_\alpha)^M$.  The
eigenvalues $\lambda_\alpha$ thus determine the thermodynamics of the
system. As was shown by Onsager and Kaufmann \cite{OandK}, the
$2^N\times 2^N$ matrix ${\cal P}$ is a spinor representation of a set
of plane rotations in $2N\times 2N$ dimensional space. The eigenvalues
of ${\cal P}$ are uniquely determined by the eigenvalues of the
corresponding plane rotation matrix $P$, which is
\begin{eqnarray}
P = \left[ \begin{array}{cccc} 
\fbox{A} & \begin{array}{cc} 0 & 0 \\ 0 & 0 \end{array} & 
\begin{array}{c} \cdots \\ \cdots \end{array} & \\
\begin{array}{cc} 0 & 0 \\ 0 & 0 \end{array} & \fbox{A}  & & \\
\begin{array}{cc} \vdots & \vdots \end{array} & & \ddots & \\
& & & \fbox{A} \end{array} \right]
\left[ \begin{array}{ccccc} 
B_{22} & \begin{array}{cc} 0 & 0 \end{array} & 
\cdots & & B_{21} \\
\begin{array}{c} 0 \\ 0 \end{array} & \fbox{B}  & & & \\
\vdots & & \fbox{B} & & \\
& & & \ddots & \\
B_{12} & & & & B_{11} \end{array} \right]  \nonumber
\end{eqnarray} 
where
\begin{eqnarray} 
\fbox{A} = \left[ \begin{array}{cc}
\cosh \theta & i \sinh \theta \\ -i \sinh \theta & \cosh \theta 
\end{array} \right] &;& \fbox{B} =  
\left[ \begin{array}{cc}
\cosh \phi & i \sinh \phi \\ -i \sinh \phi & \cosh \phi 
\end{array} \right]  \nonumber \\ 
\coth \theta &=& \cosh \phi \label{AandB}
\end{eqnarray} 
and $\phi = 2\epsilon / kT$. 

The form of the matrix $P$ suggests immediately a quantum-network
analogue.  The matrices $A$ and $B$ can be interpreted to describe
unitary evolution of a two-state or two-mode system. The matrix $P$ is
then the evolution operator over a period in the network of Fig.1.
The inputs of the network are mixed pairwise according to the
transformation $A$, and then the pairs are let to interact with the
neighbouring ones by applying the shifted set of operations $B$. By
repeating this $M$ times, a $2N\times 2M$ dimensional network can be
constructed \cite{size}.  Since $P$ contains all the physical
information of the Ising model, e.g.\ the phase transitions, we may
expect analogous phenomena in the quantum network described by
$P$. Note that our aim is {\it not} to consider a set of quantized
spin-systems at a finite temperature, like in the context of NMR
quantum computing \cite{NMR}.  Instead, we borrow from the Ising model
the abstract structure describing {\it classical statistical}
behaviour, and ask what kind of {\it quantum} behaviour it could
describe.

The physical realizations of the quantum network $P$ can be divided
into two groups. When the angle $\phi$ is real, $A$ and $B$ are
SU(1,1)-type matrices describing energy-consuming (active) operations.
Imaginary $\phi$ leads to SU(2) matrices, which correspond to
energy-preserving (passive) manipulations of the two modes or two
states. Parametric amplifiers, four-wave mixers and phase-conjugating
mirrors are SU(1,1) devices which can operate also in the quantum
regime \cite{paramp}: they could be used to build a network of active
(quantum) optical components. The corresponding passive networks could
be realized, for example, with beam splitters or fibre couplers.  Also
a network of intersecting energy levels can be described by a network
of the type in Fig.1: the avoided crossings between the levels are
identified with the operations $A$ and $B$. Corresponding physical
systems are for instance Rydberg atoms \cite{rydat} and longitudinal
electro-magnetic modes in a cavity \cite{Bow}. One can also consider
the matrix $P$ as a set of operations done by a quantum computer
\cite{Qcompexp}. For simplicity, in the following we call the SU(1,1)
components amplifiers and the SU(2) components beam splitters, but
actually mean any of the possible realizations.

Note that according to (\ref{AandB}) $A\rightarrow I_2$ when the angle
$\phi \rightarrow \infty$, and $B\rightarrow I_2$ when $\phi
\rightarrow 0$.  That is, in both of these limits the network
decomposes into sets of non-interacting modes. Thus $P$ describes a
quantum network where only nearest neighbours interact, and where a
single parameter $\phi$ determines the relative importance of the
interactions, i.e.\ the network character of the system.

The transitions in the network are determined by the eigenvalues of
$P$. The only problem in diagonalizing $P$ is the relative {\it shift}
between the sets of $A$ and $B$. This can be solved by the
discrete Fourier transform $(F_N)_{kl}= \exp (i2\pi kl/N)/\sqrt{N}
\equiv \omega^{kl}/\sqrt{N}$, because the Fourier transform of the
shift matrix $(S_N)_{kl} = \delta_{k+1,l} + \delta_{kN}\delta_{l1}$ is
diagonal: $F^{\dagger}_N S_N F_N = D_{\omega}$, where $(D_\omega)_{kl}
= \omega^{k-1}\delta_{kl}$. The whole network matrix $P$ thus
decomposes into
\begin{eqnarray}
P = I_2 \otimes F_N \left[ \begin{array}{ccc}
\fbox{$K_0$} & 0_2 & \cdots \\
0_2 & \ddots & \\ \vdots & & \fbox{$K_{N-1}$} \end{array} \right] 
I_2 \otimes F^{\dagger}_N  ,  \label{decomposed}
\end{eqnarray}
where 
\begin{eqnarray}
&&\fbox{$K_n$} = \label{K_n} \\
&&\left[ \begin{array}{cc} 
C(\theta) C(\phi) + S(\theta) S(\phi) \omega^{-n} &
C(\theta) S(\phi) - C(\phi) S(\theta) \omega^n \\
- C(\theta) S(\phi) + C(\phi) S(\theta) \omega^{-n} &
C(\theta) C(\phi) + S(\theta) S(\phi) \omega^n
\end{array} \right]   \nonumber
\end{eqnarray}
and $C\equiv \cosh$ and $S \equiv i \sinh$.  Most textbooks present the
solution of the Ising model in a slightly different form, but we have
formulated the problem as in (\ref{decomposed}) in order to make a
connection to interferometers. The usual Mach-Zehnder interferometer
affects the input states by a unitary transformation $U_{M-Z}$ which
can be formally written as
\begin{eqnarray}
U_{M-Z} = I_1 \otimes F_2 \left[ \begin{array}{cc}
K_0 & 0 \\ 0 & K_1 \end{array} \right] I_1 \otimes F_2^\dagger ,
\end{eqnarray}
where $K_n=\exp(i\phi n)$ is determined by a chosen phase $\phi$.
Thus the Ising-type network we consider
acts like an $N$-dimensional interferometer where, instead of one-mode
phase shifts, two-mode rotations are performed in between the
$N$-dimensional mixers $F_N$ and $F^\dagger_N$ \cite{ours}.

From (\ref{K_n}) one obtains the eigenvalues $e^{\pm \gamma_n}$, where
$\gamma_n$ are determined via
\begin{eqnarray}
\cosh \gamma_n &=& \cosh \theta \cosh \phi - \cos \left( \frac{2\pi
n}{N}\right) \sinh \theta \sinh \phi \nonumber \\ 
&=& \coth \phi \cosh \phi 
-  \cos \left( \frac{2\pi n}{N}\right) . 
\end{eqnarray}
An explicit expression for $\gamma_n$ is given via the
integral representation \cite{OandK}
\begin{eqnarray}
\gamma_n = \int_0^\pi \frac{d\nu}{\pi} \log \left[ 2 \left( 
\coth \phi \cosh \phi - \cos \left( \frac{2\pi n}{N}\right) 
- \cos \nu \right) \right]  \nonumber
\end{eqnarray}
As can be seen from above, the zeroth eigenvalue ($n=0$) is not a
smooth function of $\phi$: at $\gamma_0 = 0$, i.e.\ when $\cosh \phi =
\sqrt{2}$, its derivative $\frac{d\gamma_0}{d\phi}$ has a
discontinuity. In the Ising model this gives the transition temperature
$kT_c = 2.269 \epsilon$. We are now at the point to interpret what
this mathematical behaviour means physically in the case of quantum
networks. We will first consider active SU(1,1) networks, then the
passive SU(2) ones.

For the active SU(1,1) networks the amplification $G$ of the single
components $A$ and $B$ is $\cosh ^2\theta$ and $\cosh ^2 \phi$,
respectively. The critical amplification $G_c = \cosh ^2 \theta_c =
\cosh^2 \phi_c = 2$ has an interesting physical interpretation. It has
been shown \cite{Fribergetc} that for parametric amplifiers $G=2$ sets
a borderline between quantum and classical performance of the device.
For $G\ge 2$ an initially squeezed input loses the squeezing, i.e.\
its non-classical properties in the process of amplification. This is
sometimes called the "magic cloning limit"; if it did not exist, one
could reproduce quantum states, which would simply violate quantum
mechanics. By considering what happens at the critical amplification
point, as well as below and above it, we can show that $G_c$
coinciding with the "magic cloning limit" is not a mere coincidence.

In the Ising model the transition point divides regimes of order and
disorder. To see whether $G_c$ imposes any such boundary we consider
again the matrices $K_n$, now written in the form
\begin{eqnarray}
K_n = \cosh \phi \left[ \begin{array}{cc}
\coth \phi - \frac{\omega^{-n}}{\cosh \phi}
& i - \frac{i\omega^n}{\sinh \phi} \\ - i + \frac{i\omega^{-n}}{\sinh \phi} &
\coth \phi - \frac{\omega^n}{\cosh \phi} \end{array} \right]  .
\end{eqnarray}
In the limit $\phi \rightarrow 0$ the matrices $K_n$ are clearly
functions of $\omega^n$, but when $\phi \rightarrow \infty$, they
become increasingly independent of $n$; actually $G < 2$ corresponds
to the condition $|\omega^n| > |\sinh \phi |$. Considering 
Eq.(\ref{decomposed}) one sees that when the input is any vector
(specified by $n$) of the Fourier transform matrix, i.e.\ of the type
$1/\sqrt{N}[1,1,\omega^n,\omega^n,\omega^{2n},\omega^{2n}...]$, 
it will be affected by the
corresponding $K_n$ matrix. For $G>>2$, however, the $K_n$ tend to be
independent of $n$; this means that the network gives the same
response independent of the relative phases $\omega^n$ of the input modes. 
The opposite is true for $G<2$. Thus the network behaves like a phase
sensitive, i.e.\ a quantum device below $G_c=2$, and classically above
it: it is logical that $G_c=2$ coincides with the quantum-classical
border of the individual components.

Let us now consider the passive SU(2) networks 
\cite{automata}. The analogy to the
Ising model is then not one-to-one. For example (\ref{AandB}) is true
only for one trivial choice for the, now imaginary, angle $\phi$.  We
can, however, define a network which has the same basic properties as
the active ones: nodes connected by nearest neighbour interactions,
with one parameter $\phi$ quantifying the importance of these
interactions. Let us, for example, fix $\theta = i \pi/4$ and denote
$\phi' = - i \phi$. In the limits $\phi' \rightarrow 0$ and $\phi'
\rightarrow \pi/2$ the network decomposes into sets of independent
modes, while intermediate values of $\phi'$ describe a network of
interacting modes. The zeroth eigenvalue of the matrix $P$ is defined
by the equation $\cos \gamma_0 = \frac{1}{\sqrt{2}} (\cos \phi' + \sin
\phi')$. It reaches the value zero for $\phi'_c = \pi/4$, and its
derivative with respect to $\phi'$ has a singularity at this
point. The transition thus takes place at the point when all the beam
splitters are half-transmitting. Here it is, however, not a transition
between quantum and classical regimes like in the case of active
networks.  Names for the two regimes separated by $\phi'_c$ can be given
by considering a network of avoided crossings. Imagine that in Fig.1
the individual elements are actually avoided crossings between
intersecting energy levels.  At each crossing the system can either
follow the energy level adiabatically, or make a Landau-Zener
transition to the neighbouring level, that is, to show diabatic
behaviour. Indicating which of these processes is more likely, we call
the regime $\phi' < \pi/4$ diabatic and $\phi' > \pi/4$
adiabatic. Thus we have shown that the system does not evolve smoothly
from the adiabatic to the diabatic regime and vice versa, but exhibits
at the critical transmittance $t_c=\frac{1}{2}$ a transition which is
associated with singularities in the global parameters of the network.

We have now identified the regimes of behaviour separated by the
transition, both for the SU(1,1) and SU(2) networks. The essential feature
is the appearance of singularities at the transition point. In a
normal situation the tuning of local parameters, that is the
parameters of the network nodes such as beam splitters or amplifiers,
leads to a smooth change in the global properties of the network. At
the transition point this is not true. To demonstrate the
observability of this phenomenon we now consider the response
of the network in the case of two generic types of input states: the
equal superposition of all modes, and the eigenstate of one mode.

The output of the network is given by the transformation $P^M$. 
The $M$th powers of $K_n$ in Eq.(\ref{decomposed}) can be obtained
using the diagonalized form of (\ref{K_n}):
$(K_n^M)_{11} = \cosh M\gamma_n + i \sin(2\pi n/N) \sinh M\gamma_n
/\sinh \gamma_n = (K_n^M)_{22}^*$ and $(K_n^M)_{21} = (K_n)_{21} 
\sinh M\gamma_n/\sinh \gamma_n = (K_n^M)_{12}^*$.
An input in an equal superposition state with a phase periodicity
determined by $n$, that is,
$[1,0,\omega^n,0,\omega^{2n},...]/\sqrt{N}$, will be transformed into
the output state $[(K_n^M)_{11}, (K_n^M)_{12}, \omega^n (K_n^M)_{11},
\omega^n (K_n^M)_{12},...]/\sqrt{N}$. By choosing
$n$ one can thus control constructive and destructive interference in
the network ($\gamma_{n+1}>\gamma_n$). The choice $n=0$ is a special
one: at $G_c$ $\gamma_0 = 0$ and the whole network becomes
transparent. The transparency remains true independent of $M$; one may
consider this to be analogous to the appearance of long range
correlations at $T_c$ in the Ising model. Furthermore, for $n=0$ the
transition is clearly manifested in the measurable output intensities.
The global amplification coefficient is then simply $M\gamma_0$, and
its rate of change with respect to the local amplification coefficient
$\phi$, $M\frac{d\gamma_0}{d\phi}$, has a discontinuity at $G_c$.  For
an input in the eigenstate of the $2n+1$th mode, i.e.\ $[0,0,...,1,
...,0]$, the $2n+1$th output amplitude has the form $\frac{1}{N}
\sum_{m=0}^{N-1} \cosh M\gamma_m + i \sin(2\pi m/N) \sinh M\gamma_m
/\sinh \gamma_m$. Since this sum contains all
$\gamma_m$, the singularity is smoothed.  By choosing the type of
the input one can thus modify the manifestations of the transition in the
output.  Similar considerations can be carried out to find the
fingerprints of the transition in the case of SU(2) networks.

Note that due to the non-smoothness of $\gamma_0$ operations defined
by $G_c$ or $t_c$ may be unstable ones.  This has interesting
consequences: for instance, some operations to be performed by the
proposed quantum computer may hit an unstable transition point of the
whole system. Also quantum correlations and entanglement are expected
to show special behaviour near the transition.  Furthermore, it could
be illuminating to consider higher dimensional quantum interference
experiments in connection with the Ising model.  The theoretical
description of networks of avoided crossings
\cite{Harmin1,Harmin2,Bow} resembles the one presented here, with the
choice $A=B$ and with the addition of extra phase factors due to free
evolution of the system between the avoided crossings.  However, when
the free evolution phase factors are multiples of $\pi/N$, the system
reduces to the one described here, up to trivial differences. It is
interesting to note, that experimentally observed {\it recurrence
phenomena} appear in these networks exactly for such values of the
free evolution phase, that is, when the system reduces to an
Ising-type network. Moreover, approximate analytical results
predicting the recurrences are possible to derive in the adiabatic and
diabatic limits \cite{Harmin2}, but not for the case of 50:50
crossings which according to our results is the transition point
between these regimes.

In summary, we have taken a novel point of view to the networks used
in quantum computation, communication and interference experiments: we
have shown that they can exhibit transition phenomena.  Although the
individual quantum components at the network nodes are smoothly
behaving, in certain network configurations --- like the 2-D Ising
model analogue considered here --- they show non-smooth global
behaviour. For networks based on active SU(1,1) components the
transition is particularly interesting since the critical
amplification $G_c = 2$ coincides with the 'magic cloning limit' above
which an initially squeezed input loses its non-classical properties
in the process of amplification. We have shown that the whole network
too has regimes of quantum and classical behaviour: below the
transition point $G_c=2$ the network is sensitive to the phase
information in the input, for higher values of $G$ the output depends
mainly on the intensities of the inputs.  In the case of passive SU(2)
networks the critical transmittance $t_c=\frac{1}{2}$ separates
regions of adiabatic and diabatic behaviour. We have indicated how
the transition is reflected in observable quantities.

{\it Acknowledgements} We thank Prof.\ B.\ Kaufmann, Prof.\ W.\ P.\
Schleich, Dr.\ I.\ Marzoli and Dr.\ D.\ Bouwmeester for interesting
discussions, and Prof.\ P.\ Zoller for reading the manuscript and for
useful comments.

\end{multicols}


\begin{references}
\bibitem{size}
The networks should be either large enough, or imposed to periodic
boundary conditions. In the case of optical components the latter can
be easily realized.
\end{references}
\end{document}